\documentclass[11pt]{article}

\usepackage{amssymb}
\usepackage{latexsym}
\usepackage{amsmath}

\newcommand{\C}{{\mathbb C}}
\newcommand{\Cl}{{\mathbb{C}\ell}}
\newcommand{\End}{{\rm End}\ }
\newcommand{\g}{{\gamma}}
\newcommand{\G}{{\Gamma}}

\newcommand{\R}{{\mathbb R}}

\newcommand{\s}{{\sigma}}

\newcommand{\one}{{1\kern-2.5pt \text{l}} }
\newcommand{\cg}{{\cal G}}
\newcommand{\ca}{{\cal A}}

\newcommand{\cll}{{\cal L}}
\newcommand{\cm}{{\cal M}}
\newcommand{\Spin}{{\rm Spin}}

\numberwithin{equation}{section}

\begin{document}

\begin{center}
{\Large\bf THE POSSIBLE ROLE OF PURE SPINORS IN SOME\\[0.5ex]
SECTORS OF PARTICLE PHYSICS}\\
\vspace{1cm}
{\large Paolo Budinich}\\
The Abdus Salam International Centre for Theoretical Physics,\\
Trieste, Italy\\
E-mail: budinich@ictp.trieste.it; fit@ictp.trieste.it
\end{center}
\vspace{1.5cm}

\centerline{\bf Abstract}
\bigskip

It is shown how, the spinor field equation in ten dimensional
space-time: $M= \R^{1,9}$, restricted to ordinary space-time
$M=\R^{1,3}$, identifies with Cartan's equation in momentum space
$P=\R^{1,9}$, after interpreting the first four moments as
Poincar\'e translations: $P_\mu\to i\frac{\partial}{\partial
x_\mu}$. Adopting the geometry of ``simple'' spinors discovered by
\'E. Cartan \cite{one} (renamed ``pure'' by Chevalley \cite{two})
a number of results, relevant for the physics of fermions are
found. Among these, several derive naturally from Clifford
algebras and from the division algebras with which they are
correlated. Precisely the $U(1)$, at the origin of charges, derive
from complex numbers, explaining also the origin of the
charged-neutron fermion doublets, often present in nature. The
internal symmetry $SU(3)$ derives from the isomorphism $\Spin
(1,9)\simeq SL(2,{\bf o})$ where ${\bf o}$ stands for octonions
(the isomorphism is meant restricted to the Lie algebra).
``Dimensional reduction'' in this approach consists in reducing to
one half the dimensions of spinor space and to reduce by two the
ones of momentum space and consequently decoupling the equations
of motions. In this way $SU(2)$ isotopic spin symmetry is found
for the nucleon doublet interacting with the pseudoscalar pion
while the electroweak model naturally emerges with $SU(2)_L$
groups of symmetry. Both $SU(2)$ derive from quaternions since in
fact $\Cl(1,7)=H(8)$ where $H$ stands for quaternions. However
$SU(2)$ isospin does not appear as the covering of $SO(3)$ but it
derives from reflections in spinor space (which applies also to
$SU(3)$ flavour), suggesting then the redundance of the extra
dimensions when extending ordinary space-time to the ten
dimensional one: $M= \R^{1,9}$. Dimensional reduction may also
explain the origin of the 3 families of baryons and of leptons
again originating from quaternions, which also appear at the
origin of the space-time lorentzian signature. The model foresees
also the existence of 3 families of chargeless leptons: Majorana
and Weyl spinor; it could then possibly explain the origin of dark
matter.

>From ``simplicity'' instead one may derive the geometrical
structure of the momentum space where the physics of fermions is
formulated. In fact if, following Cartan, ordinary vectors are
bilinear in simple spinors, the momentum space results compact and
it consists of spheres imbedded in each other, where the quantized
theory will be free from ultraviolet divergences. Furthermore, the
spheres define invariant masses increasing with the dimensions of
the spaces and of the multiplets. Geometrical arguments are given
on why also the charges (from electric to strong) should increase
with dimensions. The possible relevance of spinor geometry and of
simplicity for several other subjects as boson field equations,
strings and superstrings are mentioned.

\section{INTRODUCTION}

In order to interpret the so-called internal symmetry phenomena,
discovered in particle physics, it has been supposed that
space-time $M$, where these phenomena evolve, may be ten
dimensional with lorentzian signature:  $M= \R^{1,9}$. In order to
obtain ordinary, four dimensional, space-time, the extra
dimensions $(>4)$ are subsequently eliminated through
``dimensional reduction''; that is by confining them in compact
manifolds of very small, unobservable size. We will tentatively
adopt this, by now almost universally adopted, hypothesis and
start by dealing, in its framework, with the dynamics of the
elementary constituents of matter, that is of fermions: the quanta
of spinor fields (bosons may be represented bilinearly, in
general, in terms of fermions).

Spinors are vectors of the automorphism space of Clifford
algebras, now these happen to be narrowly correlated with the four
existing division algebras: real and complex fields of numbers,
quaternions and octonions, and, as we will see, it is through
Clifford algebras and through the associated spinor spaces that
division algebras manifest their role at the origin of the
$SU(3)\otimes  SU(2)_L\otimes U(1)$ groups of the standard model,
at present the most appropriate to represent internal symmetry. We
will show, in fact, that they derive from the complex division
algebras. Precisely $U(1)$ from complex numbers, $SU(2)$ from
quaternions, and $SU(3)$ from octonions.

In order to prove this correlation we will adopt the spinor
geometry as formulated by its discoverer \'E. Cartan \cite{one},
who has specially stressed the great mathematical elegance of the
spinors he named ``simple'' (renamed ``pure'' by C. Chevalley
\cite{two}). We will show in fact how the Cartan's equation
defining simple or pure spinors may be identified as the spinor
field equations in the ten dimensional space-time $M$, after
setting to zero the coordinates of the extra-dimensions $(>4)$. In
this way most of the equations, adopted traditionally ad hoc by
theoretical physics, in order to describe the behaviour of fermion
multiplets, presenting internal symmetry, are naturally found. The
fact that they may be obtained from the traditional, ten
dimensional, approach after setting to zero the extra dimensions
could mean that these are redundant, at least for some of the
internal symmetry groups (like isospin $SU(2)$ and flavour
$SU(3)$). In fact what appears in the spinor field equations are
not the covering of rotation groups, like $SU(2)$ covering of
$SO(3)$, but rather the Lie algebras ($su(2)$) represented by the
generators of the Clifford algebra, meaning then that the
corresponding symmetry is generated by reflection groups in spinor
space. In this way the origin of the equations of motion can be
identified in the Cartan's equation defining spinor, and then
internal symmetry may be directly correlated with the
representation of Clifford algebra in terms of the division
algebras. So, for example, isospin $SU(2)$ derives from the fact
that  $\Cl(7,1) = H(8)$ where $H$ stands for quaternions,
$SU(3)$-flavour from $\Spin (9,1) \cong SL(2,o)$, where $o$ stands
for octonions, and the isomorphism is only meant restricted to the
Lie algebra \cite{three}.

Several more properties of Clifford algebras and of associated
spinors may plainly explain some problematic features of
elementary particles, like the origin of strong and electro weak
charges, and perhaps also of their relative values, as well as
the correlation of the 3 families of leptons, with the 3 imaginary
units of quaternions \cite{four}.

Perhaps the most remarkable feature of this approach is that, if
one adopts the elegant geometry of simple or pure spinors, the
vectors of momentum spaces (bilinear in spinors), where the spinor
field equations are formulated, are null and then the
corresponding spaces result compact, where then quantum field
theory will result, a priori, free from ultraviolet divergencies.
Furthermore, if we adopt the \'E. Cartan's conjecture that the
geometry  of simple or pure spinor might be conceived as the
elementary constituent of euclidean geometry, a striking
parallelism may emerge between geometry and physics, in so far
while euclidean geometry is well appropriate to describe classical
mechanics of macroscopic matter, the (quantum) mechanics of its
elementary constituents; the fermions, will have to be described
with the elementary constituent of euclidean geometry: the one of
simple or pure spinors. In this framework it appears obvious not
only why some elementary particle properties may already appear in
Clifford algebras of which spinors build up the representation
spaces, but also why euclidean geometry well appropriate to
represent celestial mechanics may not be any more adopted for the
description of the microscopic world of matter constituents where
then the purely euclidean concept of point -- event might have to
be substituted by that of strings: continuous sums of
null-vectors.

\section{CLIFFORD ALGEBRAS, SPINORS AND CARTAN'S EQUATIONS}

We will summarize here some elements of Clifford algebras and
associated spinor geometry necessary for the following. For more
on the subject see refs. \cite{one,two,five}.

\subsection{Hints on Spinors}

Given $V=\C^{2n}$ and the corresponding Clifford algebra $\Cl
(2n)$, generated by  $\gamma_a$; a spinor $\psi$ is a
$2^n$-dimensional vector of the endomorphism space $S$ of
$\Cl(2n): \psi\in S$ and $\Cl (2n) =\End S$.

The Cartan's equation defining $\psi$ is:
\begin{equation}
z_a\g^a\psi =0;\qquad a=1,2,\dots 2n
\end{equation}
where  $z\in V$, given $\psi\not=0 $ (implying $z_az^a=0$),
defines the $d$-dimensional totally null, projective plane
$T_d(\psi )$. For $d = n$ (maximal), $\psi$ was named simple (by
Cartan \cite{one}) or pure (by Chevalley \cite{two}), and,
according to Cartan, it is isomorphic (up to a sign) to $T_n(\psi
):= M(\psi )$.

Given $\Cl(2n) = \End S$ and $\psi ,\phi\in
S$, we have~\cite{six}:
\begin{equation}
\phi \otimes B \psi = \sum\limits_{j=0}^n F_j
\end{equation}
where $B$ is the main automorphism of $\Cl(2n)$ \cite{five} and:
\begin{equation}
 F_j =
{}_[{}\raisebox{0.5ex}{$\g_{a_1}
\g_{a_2}
\cdots
\g_{a_j}$}{}_] T^{a_1 a_2\dots  a_j} ,
\end{equation}
in which the $\gamma$
matrices are antisymmetrized and the emisymmetric tensor $T$ is
given by:
\begin{equation}
T_{a_1a_2\dots a_j}=\frac{1}{2^n}\langle B\psi
,{}_[{}\raisebox{0.5ex}{$\g_{a_1} \g_{a_2} \cdots \g_{a_j}$}{}_]
\phi\rangle\ .
\end{equation}
\bigskip

\noindent{\bf Proposition 1:} Take $\phi =\psi$ in (2.2), then
$\phi$ is simple or pure iff:
\begin{equation}
  F_0= F_1=\dots  F_{n-1} = 0;\ \  F_n\not= 0
\end{equation}
while eq.(2.2) becomes:
\begin{equation}
\phi \otimes B \phi = F_n
\end{equation}
representing the maximal totally null, plane $M(\phi )$, to which,
$\phi$ is isomorphic (up to a sign) \cite{six}.

For $\Cl(2n)$ with generators $\g_1,\g_2\dots \g_{2n}$ and with
volume element (normalized to one) $\g_{2n+1}:= \g_1\g_2\dots
\g_{2n}$ the spinors: $\psi_\pm =\frac{1}{2} (1\pm\g_{2n+1})\psi $
are named Weyl spinors, which are $2^{n-1}$-dimensional vectors of
the automorphism space of the non simple, even subalgebra $\Cl_0
(2n)$ of the simple algebra $\Cl(2n)$. Weyl spinors may be simple
or pure, and they are all so for $n\leq 3$. For $n>3$ the number
of constraint equations given by eqs.(2.5) is $1,10,66,364$ for
$n=4,5,6,7$, respectively.

\subsection{Isomorphisms of Clifford Algebras and of Spinors}

There are the isomorphisms of Clifford algebras:
\begin{equation} \Cl(2n)\simeq
\Cl_0(2n+1)
\end{equation}
both central simple. The spinors associated with $\Cl(2n)$ and $\Cl_0(2n+1)$
are named Dirac and Pauli spinors and are indicated with $\psi_D$ and $\psi_P$
respectively.

There are also the ismorphisms:
\begin{equation}
\Cl(2n+1)\simeq \Cl_0(2n+2)
\end{equation}
both non simple. The spinors associated with $\Cl_0(2n+2)$ are
named Weyl spinors and are indicated with $\psi_W$.

>From the above we get the embeddings of Clifford algebras:
\begin{equation}
\Cl(2n)\simeq
\Cl_0(2n+1)\hookrightarrow\Cl(2n+1)
\simeq\Cl_0(2n+2)\hookrightarrow\Cl(2n+2)
\end{equation}
and the corresponding ones for spinors:
\begin{equation}
\psi_D\simeq\psi_P\hookrightarrow\psi_P\oplus\psi_P
\simeq\psi_W\oplus\psi_W\simeq\Psi_D\simeq\psi_D\oplus\psi_D
\end{equation}
which implies
that a Dirac or Pauli spinor is isomorphic to a doublet of Dirac,
Pauli or Weyl spinors. These isomorphisms may be explicitly
represented. In fact let $\g_a$  be the generators of $\Cl(2n)$
and $\G_A$ those of $\Cl(2n+2)$. Then we have:
\begin{equation}
{\rm and\ for}\quad \G^{(j)}_a=\s_j\otimes\g_a:
\Psi^{(j)}=\begin{pmatrix} \psi_1^{(j)}\cr
\psi_2^{(j)}\end{pmatrix} ,\quad j=0,1,2,3
\end{equation}
where $\sigma_j$ are Pauli matrices for $j=1,2,3$, while $\sigma_0:=1_2$, and
$\Psi^{(0)}$ is a doublet of Dirac spinors while $\Psi^{(j)}$ a doublet of Weyl
$(j=1,2)$ or Pauli
$(j=3)$ spinors.

Now define the chiral projectors:
\begin{equation}
L:=\frac{1}{2} (1+\g_{2n+1});\ \ R:=\frac{1}{2} (1-\g_{2n+1}),
\end{equation}
there are the unitary transformations $U_j$:
\begin{equation}
U_j=1\otimes L+\s_j\otimes R=U^{-1}_j;\quad j=0,1,2,3
.\end{equation}
We have \cite{six}:\\ \\
{\bf Proposition 2:} Dirac, Pauli, Weyl spinor doublets are
isomorphic.

In fact:
\begin{equation}
U_j\G^{(0)}_AU^{-1}_j=\G^{(j)}_A;\ \ \ U_j\Psi^{(0)}=\Psi^{(j)};\
\ \ A=1,2\dots 2n+2;\ j=0,1,2,3\ .
\end{equation}

\noindent{\bf Remark:} The doublets $\Psi ^{(j)}$  are
geometrically isomorphic, because of eqs.(2.14), however
physically inequivalent; in fact (massive) Weyl spinors, as here
defined, may not represent free, stable particles, while Dirac and
Pauli spinors may.

\subsection{Cartan's Equations}

Cartan's equations may be obtained if we multiply eq.(2.3) from
the left by  $\gamma_a$  and from the right by $\gamma_a\phi$  and
set it to zero after summing over $a$:
\begin{equation}
\gamma_a\phi\otimes B\psi\ \gamma^a\phi = z_a\gamma^a\phi =0
\end{equation}
where
\begin{equation}
z_a=\frac{1}{2^n} \langle B\psi ,\gamma_a\phi\rangle .
\end{equation}
We have \cite{six}:\\ \\
{\bf Proposition 3:} For arbitrary $\psi$ in eq.(2.16), $z_az^a=0$ if and only
if $\phi$ is simple or pure.

Following Cartan \cite{one} we know
that for $\Cl(2n)$ only Weyl spinors, or semispinors, how he
called them, may be simple. These are spinors  with only $2^{n-1}$
non null components of the even subalgebra $\Cl_0(2n)$. Therefore
the equations for the $2^n$ component simple or pure spinors
$\phi$ may be only obtained from $\Cl(2n+2)$. Let us therefore
consider $\Cl(2n+2) = \End S$ with generators $\G_A$ and volume
element $\G_{2n+3}$ and let $\Phi$ represent a $2^{n+1}$ component
Dirac spinor. The Cartan's equation for simple or pure spinors may
be written in the form of the Weyl equation for $\Phi$:
\begin{equation}
z_A\G^A(1\pm\G_{2n+3})\Phi =0,\qquad {\scriptstyle{A}}=1,2,\dots
2n+2
\end{equation}
and
\begin{equation}
\Phi =\begin{pmatrix} \phi_+\\ \phi_-\end{pmatrix}
\end{equation}
where $\phi_\pm$ are $2^n$ component Weyl spinors associated with
$\Cl_0(2n+2)$. Adopting for $\G_A$ the representation:
\begin{equation}
\G_a=\sigma_1\otimes\g_a;\ \ \G_{2n+1}=i\sigma_1\otimes
\g_{2n+1};\ \ \G_{2n+2}=i\sigma_2\otimes 1,\ \ \  a=1,2,\dots 2n
\end{equation}
where $\g_a$ and $\g_{2n+1}$ are the generators and volume element
of $\Cl(2n)$ eqs.(2.17) become:
\begin{equation}
\left( z_a\g^a+z_{2n+1}\ \g^{2n+1}\pm z_{2n+2}\right)\phi_\pm =0
\end{equation}
where $\phi_\pm$ are simple or pure provided they satisfy the 1, 10,
66 constraint equations for $n=3,4,5$ respectively.

The above (including Proposition 3) may be easily restricted
to the real for pseudoeuclidean Clifford algebras $\Cl(1,2n-1)$ of
lorentzian signature of our interest.

\noindent   And we obtain \cite{six}:
\begin{equation}
\left( p^\pm_a\g^a+ip^\pm_{2n+1}\g_{2n+1}\pm p^\pm_{2n+2}\right)
\phi_\pm =0
\end{equation}
where the  real null vectors are obtained from (2.16) setting
$B=\G_0$; precisely:
\begin{equation}
p^\pm_A=\Phi^\dagger\G_0\G_A(1\pm\G_{2n+3})\Phi ,\ \ \
{\scriptstyle{A}}=1,2,\dots 2n+2
\end{equation}
where $\Phi^\dagger$ means $\Phi$ hermitian conjugate.

\noindent   For $\Phi\not= 0$ we must have:
\begin{equation}
p^\pm_A\ p^A_\pm =0
\end{equation}
that is $p^\pm$ must be null or optical. Because of Proposition 3
eq.(2.23) is satisfied for $p^\pm_A$ bilinear in spinors and
containing $\phi_\pm$ simple or pure. However, simplicity of
$\phi_\pm$
is not a necessary condition for the existence of solutions of
eq.(2.21), while eq.(2.23) is necessary.\\ \\

\noindent{\bf Remark 1.} For $p^\pm$ optical eq.(2.21) admits for
$\phi_\pm$ non null solutions with components functions of
$p^\pm_A$. One may also substitute the $p^\pm_A$ given in
eq.(2.22) in eq.(2.21) by which this becomes a purely spinorial
equation. However the spinors $\phi_\pm$ appearing in eq.(2.22)
have to be generally different from the ones appearing
in eq.(2.21).\\ \\
{\bf Remark 2.} Because of the isomorphism discussed in section
2.2, $\Phi$ in (2.19) could also be a doublet of Dirac or of Pauli
spinors. Precisely there are four of them $\Phi^{(j)}$ for
$j=0,1,2,3$, and they are isomorphic according to eq.(2.14).
Correspondingly there are four equivalent equations of the type
(2.20) and (2.21) for $\phi^{(j)}$ with the notation of section
2.2 where $j=0,3$ refers to Dirac and Pauli spinors respectively,
while $j=1,2$ to Weyl ones.

\section{THE EQUATIONS OF MOTION FOR FERMION MULTIPLETS, THE $U(1)$
OF THE STRONG CHARGE}

We will now apply the above formalism for spinors to the case of
the ten dimensional space-time $M=\R^{1,9}$, trying to derive the
classical equations of motion (in first quantization) to be after
quantised. For this purpose one generally starts from Lagrangians.
However dealing with spinor the transition  to the field equations
is straightforward and then we may as well start directly from the
equations of motion.

Let us then consider $\Cl(1,9) = \End S$ with generators
$\cg_\alpha$ whose Dirac spinors $\Phi\in S$ have 32 components
(they may also be Majorana since $\Cl(1,9)=\R (32)$). As seen
above we may obtain the Cartan's equations for the corresponding
simple or pure spinors starting from $\Cl(1,11)=\End S$ with
generators $\G_\beta$, volume element $\G_{13}$ and 64 components
Dirac spinor $\Xi$:
\begin{equation}
P_\beta\G^\beta (1\pm\G_{13} )\Xi =0,\ \ \ \beta = 1,2,\dots , 12
\end{equation}
where
\begin{equation}
\Xi = \begin{pmatrix} \Phi_+\\ \Phi_-\end{pmatrix}
\end{equation}
with $\Phi_\pm$ 32 component Weyl spinor. From (3.1) we obtain:
\begin{equation}
(P_\alpha\cg^\alpha +iP_{11}\cg_{11}+ P_{12})\Phi =0,\ \ \ \alpha
= 1,2,\dots ,10
\end{equation}
(that is eq.(2.22) with $n=5$) where we have taken (3.1) only with
the $+$ sign and then eliminated the super and subscripts from $P$
and $\Phi$.

Because of the isomorphisms discussed in section 2.2, we can
consider the 32-component spinor $\Phi$ as a Dirac spinor of
$\Cl(1,9)$. In fact we have only to take for the generators
$\cg_\alpha$ and volume element $\cg_{11}$ the representation:
\begin{equation}
\cg_A=1_2\otimes G_A;\ \ \cg_{9,10,11} =i\sigma_{1,2,3}\otimes
G_9,\ \ \ A=1,2,\dots ,8
\end{equation}
where $G_A$ and $G_9$ are the generators and volume element of
$\Cl(1,7)$ and:
\begin{equation}
\Phi = \begin{pmatrix} \Theta_1\\ \Theta_2\end{pmatrix}
\end{equation}
where $\Theta_1$ and $\Theta_2$ are 16 component Dirac spinors of
$\Cl(1,7)$ and eq.(3.3) becomes:
$$\begin{array}{rl}
(P_AG^A+iP_{11}G_9+P_{12})\Theta_1+i(P_9-iP_{10})G_9\Theta_2 &=
0\\
&\\
(P_AG^A-iP_{11}G_9+P_{12})\Theta_2+i(P_9+iP_{10})G_9\Theta_1 &= 0
\end{array}
\eqno(3.3')
$$

Now defining:
\begin{equation}
(P_9\pm iP_{10}) =\rho e^{i\frac{\varphi}{2}}
\end{equation}
where $\rho =\sqrt{P^2_9+P^2_{10}}$, eq.(3.3$'$) may be written in
the form:
$$
\begin{array}{rl}
(P_AG^A+iP_{11}G_9+P_{12})e^{i\frac{\varphi}{2}}\Theta_1+i\rho
G_9\Theta_2 &=
0\\
&\\
(P_AG^A-iP_{11}G_9+P_{12})\Theta_2+i\rho
G_9e^{i\frac{\varphi}{2}}\Theta_1 &= 0
\end{array}
\eqno(3.3'')
$$
which manifests the $U(1)$ invariance of the $\Theta_1$ spinor
with respect to a phase factor $e^{i\frac{\varphi}{2}}$
corresponding to a rotation of an angle $\varphi$ in the circle.
\begin{equation}
P^2_9+P^2_{10} =\rho^2
\end{equation}
generated in $\Phi$ spinor space by $\cg_9\cg_{10}$.

We will interpret it as the origin of the strong charge for the
quadruplet of fermions represented by $\Theta_1$ of which those
represented by $\Theta_2$ should then be free. Then it is natural
to conjecture that $\Theta_1:=\Theta_B$  represents baryons and
$\Theta_2:=\Theta_{\cll}$ leptons.

Before trying  to define the equations of motion for $\Theta_B$
and $\Theta_{\cll}$ let us observe that in order to have non null
solutions $\Phi$ for eq.(3.3) the vector $P$ must be null:
\begin{equation}
P_\beta P^\beta =0,\quad \beta =1,2\dots ,12
\end{equation}
and if $P_\beta$ are bilinear in $\Phi$ and if these are simple or
pure this is identically guaranteed by Proposition 3. Now
simplicity constraints for $\Phi$  implies 66 constraint equations
which should be further studied. (They forbid most transitions
between $\Theta_B$ and $\Theta_{\cll}$ and could then be at the
origin of the stability of the proton \cite{seven}.)

\section{THE BARYONS $\Theta_B$ IN $M=\R^{1,9}$  }

Let us now derive from eq.(3.3) the equation of motion for
$\Theta_B$. Reminding that $\Phi$ in eq.(3.3) is a $\Cl(1,9)$
Dirac spinor we may, exploiting the known isomorphism, adopt the
projector $\frac{1}{2}(1+\cg_{11})$, and taking into account that:
\begin{equation}
P_{11}=\Phi^\dagger\cg_0\cg_{11}(1+\cg_{11})\Phi \equiv 0\equiv
\Phi^\dagger\cg_0(1+\cg_{11})\Phi =P_{12}
\end{equation}
we easily obtain:
\begin{equation}
P_\alpha\cg^\alpha (1+\cg_{11})\Phi =
(P_AG^A+iP_9G^9+P_{10})\Theta_B =0,\ \ A=1,2,\dots ,8
\end{equation}
where $\Theta_B$ may be conceived as $\Cl(1,9)$ Weyl spinor or a
$\Cl(1,7)$ Dirac spinor, apt to represent a quadruplet of
$\Cl(1,3)$-Dirac spinors.

Observe that the 32 component spinor $\Phi$ obeying eq.(3.3) in
$P=\R^{1,11}$ has been reduced to the 16 component spinor
$\Theta_B$ obeying eq.(4.2) in $P=\R^{1,9}$; it might be conceived
as a form of ``dimensional reduction'' by which the dimension
of momentum space are decreased of two and, correspondingly, from
the equations of motion, two terms are eliminated: that is they
are decoupled; therefore in this approach dimensional reduction
consists in reducing to one half the dimension of spinor space
while decoupling two terms from the equations of motion.

\subsection{The equations in $M=\R^{1,3}$}

Let us now consider $M=\R^{1,9}$ and indicate with $x_\mu$ the
first four coordinates of ordinary space-time and with $X_j,
(j=5,6,\dots ,10)$ the other six.

Then eq.(4.2) may be considered as the Fourier transform of the
spinor field equation in $M$:
\begin{equation}
\left( i\frac{\partial}{\partial x_\mu}\ G^\mu +i\sum^{10}_{j=5}\
\frac{\partial}{\partial X_j}\ G^j\right) \Theta_B(x,X)=0
\end{equation}
where $G_{10}=1$.

Let us now restrict it to ordinary space-time $M=\R^{1,3}$ by
setting $X_j=0$. Then, defining:
\begin{equation}
i\frac{\partial}{\partial X_j}\ \Theta_B(x,X)\Big\vert_{X_j=0}
:=P_j(x)\Theta_B(x)
\end{equation}
we obtain:
\begin{equation}
\left[ i\frac{\partial}{\partial x_\mu} G^\mu +\sum^9_{j=5}\
P_j(x)G_j+P_{10}(x)\right] \Theta_B(x)=0
\end{equation}
which identifies with (4.2) if $P_\mu$ are considered as
generators of Poincar\'e translations: $P_\mu\to
i\frac{\partial}{\partial x_\mu}$ and then the $\Theta_B$ spinor
and $P_j$, for $j>5$, have to be considered $x$-dependent,
following (2.22). In eq.(4.5) the terms with $P_j$ may be
conceived as representing interactions with external fields.

We may then conclude that the Cartan's equations for $\Theta_B$
apt to represent a quadruplet of fermions presenting a strong
charge may be identified with those defined in the ten dimensional
space-time $M=\R^{1,9}$ after restricting it to $M=\R^{1,3}$.

Let us now remind that, as shown in chapter 3, that $\Theta_B$
presents a $U(1)$ invariance represented by
$e^{i\frac{\varphi}{2}}$ and from eqs.(3.6) and (4.5) we see that
the phase angle $\varphi$ results $x$-dependent, therefore to
maintain the covariance we have to introduce in eq.(4.5) a
covariant derivative with gauge potential $\ca_\mu$ and the
resulting equation (3.3) may be written in the form:
\begin{equation}
\Biggl\{\cg^\mu\left[i\frac{\partial}{\partial
x_\mu}+\frac{ig}{2}(1-\cg_9\cg_{10})\ca_\mu\right]
 +\sum^{12}_{j=1} \cg_jP_j(x)\Biggr\}\begin{pmatrix}
\Theta_B\\ \Theta_{\cll}\end{pmatrix} =0 ,
\end{equation}
where $g$ is a strong coupling contstant.

Let us see which are the further physical information which can be
directly read from eqs.(4.2) or (4.5). To admit non null solutions
for $\Theta_B$, the vector $P$ has to be null\footnote{If, in the
framework of Cartan's geometry, we express $P$ bilinearly in terms
of $\Theta_B$, eq.(4.7) is an identity, because of Proposition 3,
if $\Theta_B$ is simple and it satisfies the ten constraint
equations; which could indicate that, while these are
geometrically relevant, they are not necessary conditions for the
existence of physical fermions, which could be then represented
also by non simple or pure spinors.}:
\begin{equation}
P_\alpha P^\alpha =0\qquad \alpha =1,2,\dots ,10
\end{equation}
which defines the invariant mass $\cm$:
\begin{equation}
P_\mu P^\mu =P^2_5+P^2_6+P^2_7+P^2_8+P^2_9+P^2_{10} =\cm^2
\end{equation}
in turn defining $S^5$  presenting an $SU(6)$ group of symmetry
orthogonal to the Poincar\'eÇ group, therefore the maximal
internal symmetry for the quadruplet of fermions represented by
$\Theta_B$ will be $SU(4)$, covering of $SU(6)$. Observe that
dimensional reduction decouples the equations of motion and
therefore the invariant mass $\cm$ will be smaller for fermion
doublets and singlets.

\subsection{Flavour $SU(3)$}

We know that spinors are vectors of the representation spaces of
Clifford algebras and therefore we may expect that already in
these we may read some of their properties. Now in this case:
$\Spin (1,9)\cong SL(2,o)$ where $o$ stands for octonions, and the
isomorphism is only valid if restricted to the Lie algebra
\cite{three}. The automorphism group $G_2$ of octonions has a
subgroup $SU(3)$ if one of its seven imaginary units is fixed, and
this could explain the origin of the $SU(3)$ symmetry of flavour.
In fact it may be shown \cite{eight} that some of the generators
$\cg_\alpha$ of $\Cl(1,9)$ may be expressed in terms of the seven
imaginary units of octonions: $e_1, e_2, e_3, e_7$ and $\hat e_1,
\hat e_2, \hat e_3$, where $\hat e_n:= e_ne_7$ for $n=1,2,3$.

Precisely, after defining $e_n:=-i\sigma_2\otimes \sigma_n$ and
$e_7:i\sigma_3\otimes 1_2$ \cite{nine} one may define:
\begin{equation}
\cg_{6+n}:=e_n\otimes\G_7;\ \ \ \cg_{11}:= ie_7\otimes 1_8,\ \
n=1,2,3.
\end{equation}

The resulting algebra is non associative nevertheless it may be
closed either redefining matrix multiplications \cite{nine} or
redefining products of octonian units expressed in terms of
Clifford algebra generators \cite{eight}; however, we do not need
to deal here with this problem since we will only use the
associative $SU(3)$ subalgebra of octonions. In fact it is known
\cite{ten} that the complex octonions:
\begin{equation}
U_\pm=\frac{1}{2} (1\pm ie_7);\ \ \ V^{(n)}_\pm = \frac{1}{2}
e_n(1\pm ie_7), \ \ n=1,2,3,
\end{equation}
define an $SU(3)$ invariant algebra for which $V^{(n)}_+$ and
$V^{(n)}_-$ transform as the $(3)$ and $(\bar{3})$ representations
of $SU(3)$ respectively, while $U_+$ and $U_-$ as singlets.
Therefore, writing eq.(4.2) in the form:
\begin{equation}
P_\alpha\cg^\alpha (1+\cg_{11})\Phi = \left( P_a\cg^a+\sum^3_{n=1}
P_{6+n} \cg_{6+n}+P_{10}\right) (1+\cg_{11})\Phi =0,\ \ a=1,2\dots
,6
\end{equation}
we may through (4.9) and (4.10) insert in it $U_+= \frac{1}{2}
(1+\cg_{11})$ and $V^{(n)}_+=\frac{1}{2} \cg_{6+n}(1+\cg_{11})$:
\begin{equation}
P_\alpha\cg^\alpha (1+\cg_{11})\Phi =\left( P_a\cg^a+\sum^3_{n=1}
P_{6+n} V^{(n)}_++P_{10}\right) U_+\Phi =0
\end{equation}
which for
\begin{equation}
P_{6+n} = \Phi^\dagger \cg_0\cg_{6+n} (1+\cg_{11})\Phi
=\Phi^\dagger G_0V^{(n)}_+\Phi
\end{equation}
is $SU(3)$ covariant.

Now acting with $V^{(n)}_+$ on the Cartan's standard spinors (or
on the vacuum of a Fock representation of spinor space \cite{six})
one could represent the 3 spinors representing quarks and then the
known $3\times 3$ representation of the pseudo-octonions algebra
\cite{eleven} for baryons conceived as 3-quark states, which could
represent $SU(3)$ flavour symmetry for baryons. Observe that the
$\cg_{6+n}$ from which $SU(3)$ derives are reflection operators in
spinor space.

\subsection{Color $SU(3)$}

Complex octonions may be also defined with the first four
generators $\cg_\mu$ in eqs.(4.6) and (4.11). They are
\cite{eight}:
\begin{equation}
V^{(0)}_\mu =\frac{-i}{2} \cg^{(0)}_\mu (1+\cg_{11});\quad
V^{(n)}_\mu =\frac{-i}{2} \cg^{(n)}_\mu (1+\cg_{11}),\quad n=1,2,3
\end{equation}
where $\cg^{(0)}_\mu =e_0\otimes \G_\mu$ where $e_0=1$ and
$\cg^{(n)}_\mu = e_n\otimes \G_\mu
=i\sigma_2\otimes\sigma_n\otimes\G_\mu$ and the indices $n$ refers
to the isomorphism discussed in section 2.2. As before they
transform as singlets and triplets for $SU(3)$. We have seen that
$\Theta_B$ presents an $U(1)$ symmetry in form of a local phase
factor which will impose a covariant derivative and eq.(4.12)
takes the form \cite{eight}:
$$
P_\alpha\cg^\alpha (1+\cg_{11})\Phi =\left[
i\left(\frac{\partial}{\partial x_\mu} -gA^\mu_{(n)}\right)
V^{(n)}_\mu +P_5\cg_5+P_6\cg_6 +\sum^3_{n=1} P_{6+n}V^{(n)}_+ +
P_{10}\right] U_+\Phi =0, \eqno(4.12')
$$
where
$$
A^{(n)}_\mu = \Phi^\dagger\cg_0\cg^{(n)}_\mu (1+\cg_{11})\Phi =
2i\Phi^\dagger \cg_0 V^{(n)}_\mu \Phi
$$
and then, in the covariant derivative, an $SU(3)$ covariant term
appears interpretable as $SU(3)$-color. It could allow to
correlate the three colors with the imaginary units of quaternions
(Pauli matrices). On acting with $V^{(n)}_\mu$ on standard spinors
one would generate colored quarks and the baryons as 3 colored
quark states. If we impose that the physical fermions are
represented by Dirac spinors then: $\cg^{(0)}_\mu = e_0\otimes
\G_\mu$ and if inserted in (4.12$'$) they give $V^{(0)}_\mu$ which
are $SU(3)$-color singlets and the impossibility of physical
colored baryons could be correlated with the fact that massive
Weyl spinors may not represent free particles.

\subsection{The nucleon doublet, isospin $SU(2)$ and the $U(1)$
of the electric charge}

Let us now perform a further dimensional reduction. We will use
the projector $\frac{1}{2}(1+G_9)$ which will reduce the
16-component spinor $\Theta_B$ to an 8-component one $N$:
\begin{equation}
\frac{1}{2} (1+G_9)\Theta_B=N=\begin{pmatrix} \psi_1\\ \psi_2
\end{pmatrix}
\end{equation}
and correspondingly as easily verified:
\begin{equation}
P_9=\Theta^\dagger_BG_0G_9(1+G_9)\Theta_B\equiv 0\equiv
\Theta^\dagger_BG_0(1+G_9)\Theta_B =P_{10}
\end{equation}
and the equation of motion for $N$ will be:
\begin{equation}
\frac{1}{2} (1+G_9)\Theta_B=\left(i\frac{\partial}{\partial x_\mu}
\G^\mu+P_5\G_5+P_6\G_6+P_7\G_7+P_8\right)N=0
\end{equation}
We will now suppose that $\psi_1$ and $\psi_2$ in (4.15) represent
$\Cl(1,3)$ Dirac spinor, by which we have to impose:
\begin{equation}
\G^{(0)}_\mu = 1\otimes\g_\mu ;\quad \G^{(0)}_{5,6,7}
=-i\sigma_{1,2,3}\otimes\g_5
\end{equation}
and (4.17) becomes:
\begin{equation}
\left( i\frac{\partial}{\partial x_\mu} \g^\mu
-i\boldsymbol{\pi}\cdot\boldsymbol{\sigma}\otimes\g_5 + P_8\right)
\begin{pmatrix} p\\ n\end{pmatrix} =0
\end{equation}
where
\begin{equation}
\boldsymbol{\pi} = N^\dagger\G_0\boldsymbol{\sigma}\otimes\g_5N
\end{equation}
and where we have set $\psi_1=p$ and $\psi_2=n$ representing the
proton and neutron respectively, since eq.(4.19) is well
representing the $SU(2)$-isospin covariant nucleon-pion equation
of motion. Observe that the pseudoscalar nature of the pion
isotriplet $\boldsymbol{\pi}(x)$ derives from having imposed that
$p$ and $n$ are $\Cl(1,3)$-Dirac spinors because of which
$\G_5,\G_6$ and $\G_7$ in (4.18) have to contain $\g_5$, in order
to anticommute with $\G_\mu$.

Observe that isospin $SU(2)$ is generated by $\G_5,\G_6,\G_7$
which are reflection operators in the spinor space spanned by $N$.
It is not then the covering of $SO(3)$ in the space spanned by
$X_5,X_6,X_7$ and it could not be otherwise, since, to obtain
eq.(4.19) these coordinates have been set to zero. It derives
instead from quaternions. In fact $\Cl(1,7)=H(8)$ and consequently
eq.(4.19) is manifestly quaternionic and then one may affirm that,
according to this derivation,  quaternions appear to be at the
origin of the isospin symmetry of nuclear forces.

Let us now write explicitly (4.19) for $p$ and $n$: defining
$p_5\pm ip_6=\rho^{i\frac{\varepsilon}{2}}$, it may be easily
brought to the form:
\begin{equation}
\begin{array}{rl}
(P_\mu\g^\mu +iP_7\g_5+P_8)e^{i\frac{\varepsilon}{2}}p-\rho\g_5n&=0\\
&\\
(P_\mu\g^\mu
-ipP_7\g_5+P_8)n+\rho\g_5e^{i\frac{\varepsilon}{2}}p&=0
\end{array}
\end{equation}
where $\varepsilon$ is an angle of rotation in the circle
\begin{equation}
P^2_5+P^2_6=\rho^2
\end{equation}
generating the $U(1)$ phase invariance of the proton $p$
represented by $e^{i\frac{\varepsilon}{2}}$ and generated by
$J_{56}=\frac{1}{2}[\G_5,\G_6]$.

Since $\varepsilon (x)$ is local it imposes a covariant
derivative, and then an electric charge, $e$ for the proton: in
fact the resulting equation of motion for the doublet $N$ is:
\begin{equation}
\left\{\left[ i\frac{\partial}{\partial x_\mu} -\frac{e}{2}
(1-\G_5\G_6)A_\mu\right]\g^\mu
-i\boldsymbol{\pi}\cdot\boldsymbol{\sigma}\otimes\g_5+P_8\right\}
\begin{pmatrix} p\\ n\end{pmatrix} =0
\end{equation}
well representing the nuclear and electromagnetic properties of
the nucleon. One may also represent in one quadruplet $p, n$,
electron and neutrino and the electric charges of proton and
electron result of opposite sign \cite{seven}.

\section{THE LEPTONS $\Theta_{\cll}$}

The equation for $\Theta_{\cll}$ will be:
\begin{equation}
p_\alpha\cg^\alpha (1-\cg_{11})\Phi =0\quad \alpha =1,2,\dots 10
\end{equation}
from which we will derive the equation in $M^{1,3}$:
\begin{equation}
\left[ i\frac{\partial}{\partial x_\mu}G^\mu +\sum^9_{j=5}
p_j(x)G_j-p_{10}(x)\right]\Theta_{\cll} =0
\end{equation}

\subsection{The equations of motion for $\Theta_{\cll}$
and the $U(1)$ for the electroweak charge}

Let us now suppose $\Theta_{\cll}$ to have the form:
\begin{equation}
\Theta_{\cll} =\begin{pmatrix} L_1\\ L_2\end{pmatrix} =
\begin{pmatrix} \ell_{11}\\ \ell_{12}\\ \ell_{21}\\ \ell_{22}\end{pmatrix}
\end{equation}
where $L_1,L_2$ are $\Cl(1,7)$-Dirac spinors, and then take for
the generators $G_A$ and volume elements $G_9$ the representation:
\begin{equation}
G_a=1\otimes\G_a;\quad G_{7,8,9} =i\sigma_{1,2,3}\otimes\G_7\quad
a=1,2,\dots 6
\end{equation}
where $\G_a,\G_7$ are generators and volume element of $\Cl(1,5)$,
and we obtain, operating as above, for $L_1$ and $L_2$ the system
of equations:
\begin{equation}
\begin{array}{rl}
(p_a\G^a+ip_9\G_7-p_{10})e^{i\frac{\tau}{2}}L_1+i\rho\G_7L_2&=0\\
&\\
(p_a\G^a-ip_9\G_7-p_{10})L_2+i\rho\G_7e^{i\frac{\tau}{2}}L_1&=0
\end{array}
\end{equation}
where now $\tau$ is an angle of rotation on the circle
$$
p^2_7+p^2_8=\rho^2
$$
at the origin of the $U(1)$ phase invariance of $L_1$ in spinor
space generated by $G_7 G_8$. We will suppose it at the origin of
a charge (different from the strong one generated by $\cg_9
\cg_{10}$) for the lepton doublet $L_1$ from which instead $L_2$
should be free.

\subsection{The $SU(2)_L$ phase invariance and the electroweak model}

Let us suppose that the four leptons in $\Theta_{\cll}$  of
eq.(5.3) may represent four free fermions in space-time
$M=\R^{1,3}$; they should then be either Dirac or Pauli spinors,
and, for the representation of $G_\mu$ we have then four possible
choices, diagonal in $\g_\mu$: generators of $\Cl(1,3)$,
precisely:
\begin{equation}
\begin{array}{rl}
G^{(0,0)}_\mu = 1\otimes 1\otimes\g_\mu;\quad &G^{(0,3)}_\mu = 1
\otimes \sigma_3\otimes\g_\mu;\\
&\\
G^{(3,0)}_\mu = \sigma_3\otimes 1\otimes\g_\mu;\quad
&G^{(3,3)}_\mu = \sigma_3\otimes \sigma_3\otimes\g_\mu .
\end{array}
\end{equation}
Correspondingly we have four possible systems of equations for the
doublets $L_1$ and $L_2$. Let us take as an example
$G^{(0,3)}_\mu$, then the other generators are:
\begin{equation}
\begin{array}{rl}
G^{(0,3)}_{5,6} = i\sigma_1\otimes \sigma_{1,2}\otimes 1_4;\quad
&G^{(0,3)}_7 =i\sigma_1
\otimes \sigma_3\otimes\g_5;\\
&\\
G^{(0,3)}_8 = i\sigma_2\otimes 1_2\otimes\g_5;\quad &G^{(0,3)}_9 =
i\sigma_3\otimes 1_2\otimes\g_5
\end{array}
\end{equation}
where $\g_5$ is the volume element of $\Cl(1,3)$, and the
corresponding equations are:
\begin{equation}
\begin{array}{rl}
(p_\mu\sigma_3\otimes\g^\mu +p_9\g_5-p_{10})L_1+[
i(p_5\sigma_1+p_6\sigma_2+p_7\sigma_3\otimes\g_5)+p_8\g_5] L_2
&=0\\
&\\
(p_\mu\sigma_3\otimes\g^\mu -p_9\g_5-p_{10})L_2+[
i(p_5\sigma_1+p_6\sigma_2+p_7\sigma_3\otimes\g_5)-p_8\g_5] L_1
&=0.
\end{array}
\end{equation}

>From the other $G_\mu$ in (5.6) we obtain three other systems
which differ for the distribution of the $\g_5$. Let us now define
the projectors:
$$
{\cal L} :=\frac{1}{2}(1+1\otimes\g_5);\quad {\cal R}
:=\frac{1}{2} (1-1\otimes\g_5)
$$
and introduce the notation:
$$
{\cal L}L_1:=L_{1L};\quad {\cal R}L_1:=L_{1R}
$$
and the same for $L_2$.

We have $\g_5L_{1L}=L_{1L}$ and $\g_5L_{2L}=L_{2L}$. Applying
${\cal L}$ to (5.8), after defining
$p_5\sigma_1+p_6\sigma_2+p_7\sigma_3:=\frac{1}{2}\boldsymbol{\omega}\cdot
\boldsymbol{\sigma}$ we obtain:
$$
\begin{array}{rl}
p_\mu\sigma_3\otimes\g^\mu L_{1R} +(p_9-p_{10})L_{1L} +\left(
p_8+\frac{i}{2}\boldsymbol{\omega}\cdot\boldsymbol{\sigma}\right) L_{2L}
&=0\\
&\\
p_\mu\sigma_3\otimes\g^\mu L_{2R} -(p_9+p_{10})L_{2L} -\left(
p_8-\frac{i}{2}\boldsymbol{\omega}\cdot\boldsymbol{\sigma}\right) L_{1L}
&=0. \end{array}\eqno(5.8')
$$

Let us now define:
\begin{equation}
\left( p_8-\frac{i}{2}\boldsymbol{\omega}\cdot\boldsymbol{\sigma}\right)
=\rho e^{-\frac{i}{2}\boldsymbol{\omega}\cdot\boldsymbol{\sigma}}=\rho
e^{\frac{q}{2}}
\end{equation}
where $q=-i{\boldsymbol{\sigma}\cdot\boldsymbol{\omega}}$ is an
imaginary quaternion and $\rho = \sqrt{p^2_5+p^2_6+p^2_7+p^2_8}$.
Eq.(5.8$'$) may be brought to the form:
$$
\begin{array}{rl}
e^{\frac{q}{2}}p_\mu\sigma_3\otimes\g^\mu L_{1R} + (p_9-p_{10})
e^{\frac{q}{2}}L_{1L} +\rho L_{2L} &=0\\
&\\
p_\mu\sigma_3\otimes\g^\mu L_{2R} - (p_9+p_{10}) L_{2L} +\rho
e^{\frac{q}{2}}L_{1L} &=0
\end{array}
\eqno(5.8'')
$$
manifesting a quaternionic on $SU(2)_L$ phase invariance for
$L_{1L}$, which being $x$ dependent will induce a covariant
derivative with a non abelian gauge interaction:
\begin{equation}
D_\mu = \frac{\partial}{\partial x_\mu} -
i\boldsymbol{\sigma}\cdot\bf{W}_\mu
\end{equation}
for $L_{1L}$  only and then the Dirac equation for
$L_1=\begin{pmatrix} e\\ \nu_L\end{pmatrix}$ where $e$
represents the electron and $\nu_L$ the left-handed neutrino is easily
obtained to be:
\begin{equation}
\left(i\frac{\partial}{\partial x_\mu}\g^\mu +m\right)
\begin{pmatrix} e\\ \nu_L\end{pmatrix} +\boldsymbol{\sigma}\cdot
{\bf W}_\mu \g^\mu \begin{pmatrix} e_L\\
\nu_L\end{pmatrix} +B_\mu\g^\mu e_R =0
\end{equation}
where ${\bf W}_\mu$ is an isotriplet vector field and $B_\mu$ an
isosinglet. Eq.(5.11) is notoriously the geometrical starting
point of the electroweak model. It derives, like isospin, from the
quaternion division algebra and it may be also obtained
\cite{seven} from the isomorphisms discussed in section 2.2 (from
which it also here effectively derives).

\section{DIMENSIONAL REDUCTION AND FAMILIES FOR $\Theta_B$ AND $\Theta_{\cll}$}

Let us now return to the problem of dimensional reduction which,
as mentioned in chapter 4, consists, in our approach, in reducing
to one half the dimensions of spinor space, while decoupling two
terms from the equations of motion. In section 4.4 we have adopted
the projector $\frac{1}{2} (1+G_9)$ to reduce the 16 component
spinor $\Theta_B$ to the 8 component one $N$. Now the six first
generators $G_a$ of $\Cl(1,7)$ are traditionally set in the form:
\begin{equation}
G^{(j)}_a =\sigma_j\otimes\G_a;\quad j=0,1,2,3,\quad a=1,2,\dots
,6
\end{equation}
where $\G_a$ are the generators of $\Cl(1,5)$ and $\sigma_0=1$.
The above projector is obtained for $j=1,2$, therefore we will now
indicate $\frac{1}{2} (1+G_9)\Theta_B=\Psi^{(1,2)}$, and we have
seen in eq.(4.16) that $P_9\equiv 0\equiv P_{10}$. Instead  for
$j=0$ and $j=3$ we obtain the projectors
\begin{equation}
\frac{1}{2} (1+iG_7G_8)\Theta_B=\Psi^{(0)}\ \ \textrm{and}\ \
\frac{1}{2} (1+iG_8G_9)\Theta_B=\Psi^{(3)}
\end{equation}
respectively and, correspondingly:
$$P_7=\Theta^\dagger_BG_0G_7(1+iG_7G_8)\Theta_B
\equiv 0\equiv\Theta^\dagger_BG_0G_8(1+iG_7G_8) =P_8$$ and\hfill
(6.3)
$$P_8=\Theta^\dagger_BG_0G_8(1+iG_8G_9)\Theta_B
\equiv 0\equiv\Theta^\dagger_BG_0G_9(1+iG_8G_9) =P_9$$
respectively and the corresponding so reduced 8 dimensional space,
for $\Theta_B$ simple or pure (one constraint equation), is null
defining then the invariant masses.
\setcounter{equation}{3}
\begin{eqnarray}
P_\mu P^\mu &=& P^2_5+P^2_6+P^2_7+P^2_8 = M^2_{(1,2)} \nonumber \\
P_\mu P^\mu &=& P^2_5+P^2_6+P^2_7+P^2_{10} = M^2_{(3)}\\
P_\mu P^\mu &=& P^2_5+P^2_6+P^2_9+P^2_{10} = M^2_{(0)}\nonumber
\end{eqnarray}
Because of the isomorphisms discussed in section 2.2 the three
spinors and $\Psi^{(0)}, \Psi^{(3)}$ and $\Psi^{(1,2)}$ may
represent doublets of fermions which, in the limit of absence of
external, strongly interacting fields (the pion in eq.(4.19)),
will obey the Dirac equations.

Their masses might be different, and, since we know that, in
general, the characteristic energy of the phenomena increases with
the dimension of the $P$-space we could expect
$M_0>M_3>M_{(1,2)}$.

We can now repeat the same with the lepton quadruplet
$\Theta_{\cll}$ and we will obtain three lepton doublets
$L^{(1,2)}_1,L^{(3)}_1,L^{(0)}_1$ representing charged-neutral
fermions of which the changed partners will obey to the Dirac
equation since now the strong charge should be zero and then the
strongly interacting fields, the pions, should be absent. If these
are represented by $p_5,p_6,p_7$ (like in eq.(4.19)), then the
corresponding invariant mass equations become:
\begin{eqnarray}
p_\mu p^\mu &=& p^2_8 = m^2_{(1,2)}\nonumber \\
p_\mu p^\mu &=& p^2_{10} = m^2_{(3)}\\
p_\mu p^\mu &=& p^2_9+p^2_{10} = m^2_{(0)}\nonumber
\end{eqnarray}
and then, adopting the above hypothesis, we could conjecture that
$L^{(1,2)}_1,L^{(3)}_1,L^{(0)}_1$ represent the electron-neutrino,
muon-neutrino, and tau-neutrino doublets respectively.

We have then that the above geometrical structure foresees
three families of leptons and the corresponding ones of baryons as
discovered in nature. Their origin may be assigned to the
quaternion division algebra; in fact it derives from the existence
of the three projectors which in turn derive from the fact that
the generators $G_a$ may be set in the form (6.1) manifestly
correlated with quaternions.

The possible quaternionic origin of three lepton families was also
obtained by T. Dray and C. Monogue \cite{four} from octonion
division algebra correlated with $\Cl(1,9)$.

\section{THE CHARGELESS LEPTONS}

If we act on $\Theta_{\cll}$ with $\frac{1}{2}(1-G_9)$  or with
the other two possible projectors discussed in chapter 6 we will
obtain three families of leptons $L^{(1,2)}_2,L^{(3)}_2,L^{(0)}_2$
which, however, from what is seen above, should be free from both
strong and electroweak charges. Therefore we will need a further
dimensional reduction. We will operate it through the projector
$\frac{1}{2}(1\pm\G_7)$ and we obtain:
\begin{equation}
\frac{1}{2}(1\pm\G_7)L_2=\psi_\pm
\end{equation}
and since
\begin{equation}
p_7=L^\dagger_2\G_0\G_7(1\pm\G_7)L_2\equiv 0\equiv
L^\dagger_2\G_0(1\pm\G_7)L_2=p_8
\end{equation}
the equation will be:
\begin{equation}
(p_\mu\g^\mu +p_5\g_5\pm p_6)\psi_\pm =0
\end{equation}
Now for $\psi$ real $p_6=\tilde\psi\psi\equiv 0$ and (7.3) reduces
to:
$$ (p_\mu\g^\mu +p_5\g_5)\psi =0 \eqno(7.3')$$
which is a candidate for the Majorana equation. A further
dimensional reduction through $(1\pm\g_5 )$  reduces (7.3) to
\begin{equation}
p_\mu\g^\mu (1\pm\g_5)\psi =0
\end{equation}
since $p_5\equiv 0\equiv p_6$ that is the Weyl equation for
massless neutrinos.

Then the final reduction brings us to Majorana and Weyl spinors.
They then naturally appear as the most elementary form of
fermions. As mentioned, there should be three families of them and
they should be created at the Big Bang, and in a roughly
computable fraction, as stable, missing known interactions,
including electroweak ones and keeping only gravitational ones.
Through gravitational interactions they could also have been
created and accumulated during the life of the universe during
high energy events connected with supernovae and black holes.
Could they then contribute in furnishing an answer to the great
problem of black matter?

It is remarkable that in this picture it is not Dirac's equation
the basic equation for fermions, but rather Majorana and Weyl
ones. One could then try to start from these the construction of
both the fermion multiplets and of space-time or momentum space
and in so doing also the lorentzian signature of Minkowski
space-time (or of $\Cl(1,3)$) appears to be unambiguously defined
\cite{seven}, once more, from quaternion division algebra of which
in fact it results simply to be the image.

\section{THE BOSON FIELD EQUATIONS}

In our approach the boson fields appear at first as external
fields as was the case for the electromagnetic potential $A_\mu$
and of the pion field $\boldsymbol{\pi}$ in eq.(4.23). However,
since the boson fields here are bilinear in spinor fields also
their equations of motion should be obtainable from the spinor
field equation. In fact it was shown \cite{seven} that from Weyl
eq.(7.4) it is possible to obtain for
\begin{equation}
F^{\mu \nu}_\pm = \tilde\psi [\g^\mu , \g^\nu ] (1\pm\g_5)\psi
\end{equation}
the equations
\begin{equation}
p_\mu F^{\mu\nu}_+ =0;\quad \varepsilon_{\mu\nu\rho\tau}p^\nu
F^{\rho\tau}_- =0
\end{equation}
That is Maxwell's homogeneous equations, and also the non
homogeneous ones \cite{twelve}.

Observe that in eq.(8.1) the electromagnetic field
$F^{\mu\nu}_\pm$ is bilinear in the Weyl spinor fields, which does
not imply that, in the quantized theory, the photon should be conceived as a
bound state of neutrinos. In fact ``the neutrino theory of light''
is notoriously unviable \cite{thirteen}.

In a similar way it is possible to obtain the equations of motion
for the pion field $\boldsymbol{\pi} (x)$, not necessarily as a
bound state of proton-neutron \cite{fourteen}. It would be
reasonable to expect the possibility to extend this to the other
known boson fields.

\section{THE ROLE OF SIMPLICITY}

We have seen how the geometry of simple or pure spinors, if
correlated with division algebras, seems to be appropriate to
explain several features of elementary particle phenomenology,
like the origin of charges, of families; the origin of internal
symmetry groups: of the standard model $SU(3)\otimes SU(2)\otimes
U(1)$, in the dynamical sector of the equations, and of isospin
$SU(2)$, of favour $SU(3)$ up to a possible maximum of $SU(4)$ all
deriving from reflections, in the rest of the equations. For these
matters the simplicity constraint-equations, in number 1,10,66 for
the 8,16,32-component spinors we dealt with, do not seem to play a
critical role.

Simplicity, through  Proposition 3 in chapter 2, mainly imposes
the nullness of the momentum-spaces bilinearly constructed with
those spinors. This in turn implies that: to the embedding of
4-dimensional spinor spaces in 8-16- and 32-dimensional ones there
correspond the parallel embedding of the corresponding
4-dimensional vector spaces in 6-,8- and 10-dimensional ones,
which, being null define compact manifolds (spheres) imbedded in
each other, where the concept of infinity is absent, each space
being isomorphic (up to a sign), as discovered by \'E. Cartan, to
the spinors themselves building up geometrical structures of great
simplicity and elegance. This was presumably the reason why \'E.
Cartan named them ``simple'', despite the apparent complication of
the constraint equations.

Because of this one might expect that simplicity, while not
determinant for some features of phenomenology - where one could
then expect to have also to deal with non simple or generalized
pure spinors \cite{fifteen}  - determines instead rigorously the
underlying projective geometry and then may have deep meanings
helpful for the solution of some as yet open problems.

We will try to list and  discuss preliminarily some of these
problems and tentatively guess some possible answers suggested by
that simple geometry.

\subsection{The masses}

If momentum spaces are null or optical they define invariant
masses spheres like eq.(4.8). Before discussing their role in
physics we need to remind that the terms appearing in these
equations are representing external boson-field interactions and
we need then first to determine the equations of motion for these
boson fields. Furthermore, those terms may contain multiplicative
coupling constants, like $P_5,P_6,P_7$ in  eq.(4.17) which should
contain, as a factor, the pion-nucleon coupling constants.
Therefore this problem is correlated with the one  of determining
the values of the coupling constants and charges, after which it might present
aspects of great relevance for physics.

\subsection{The charges}

The values of the charges are traditionally taken from the
experimental data and as such inserted by hand in the equations of
motion. In the present approach instead, in which the equations of
motions are unambiguously derived from spinor geometry in momentum
space, which in turn appears to consist in spheres
imbedded in each other, one should expect that everything should
be unambiguously defined including coupling constants and nothing
but boundary conditions, should be left to be added by hand. One
suggestion that this might be true and realizable comes from the
old Fock equation \cite{sixteen} for the hydrogen atom written
precisely in the sphere $S_3$ of momentum space which may be set
in the following adimensional form:
\begin{equation}
\psi ({\bf u}) = \frac{1}{V(S_3)} \ \frac{e^2}{\hbar c} \
\frac{mc}{p_0} \int_{S_3} \frac{\psi({\bf u}')}{({\bf u}-{\bf
u}')^2} \ d^3{\bf u}'
\end{equation}
where ${\bf u}$ is a unit vector of $S_3$, $p_0$ a unit of
momentum, $m$ the mass of the electron, and where also the
electric coupling constant $e$ appears inserted in the
adimensional fine structure constant $\alpha^{-1}  = \frac{\hbar
c}{e^2} = 137,036\dots$. Eq.(9.1) manifests the $SO(4)$ covariance
induced in the two body system by the Coulombian potential and is
obtained by Fock through a harmonic analysis on $B_3\to  S_3$:
\begin{equation}
h(x)=\frac{1}{V(S_3)} \int_{S_3} G(x,{\bf u}') h({\bf u}')d^3{\bf
u}'
\end{equation}
where $x\in B_3, G(x,{\bf u}')$ is a Green function and $h(x)$ is
harmonic in $B_3$. Then for $h(x)=r^{n-1}\psi_n({\bf u})$,
spherical harmonic on $S_3$ eq.(9.1) is obtained representing the
H-atom eigenfunctions and energy levels for stationary states (and
could be also extended to the scattering states of the continuous
spectrum).

It was then conjectured\footnote{In a paper in progress  with P.
Nurowski.} that also the factor in front of eq.(9.1); that is
$\alpha$, could be obtained from a harmonic analysis from the five
dimensional classical domain of the IV type $D=Q_5$ to $D'=
e^{i\varphi} S_4$ obtaining:
\begin{equation}
h({\bf u}) = \frac{1}{V(S_3)} \ \frac{2\pi J}{V(S_4)V(Q_5)}
\int_{S_3} \frac{h({\bf u}')}{({\bf u}-{\bf u}')^2} \ d^3{\bf u}'
\end{equation}
where $J$ derives from the integration of a Jacobian.

We have seen that in our approach simplicity defines the compact
domains imbedded in each other: they are
spheres, and since in case the strong charge $g$ of the baryon
$\Theta_B$ is defined, as shown in chapter 3, from $P_\beta
\G^\beta (1\pm\G_{13})\Xi =0$ and following ones for $\Cl(1,11)$,
one can repeat the above derivation to find:
\begin{equation}
h({\bf u}) = \frac{1}{V(S_3)} \ \frac{2\pi J'}{V(S_{11})V(B_{12})}
\int_{S_3} \frac{h({\bf u}')}{({\bf u}-{\bf u}')^2} \ d^3{\bf u}'
\end{equation}
where now one could interpret the factor in front of the integral
as proportional to $g^2/\hbar c$ where $g$ is the strong charge
(also two body strong forces are Coulombian for high momenta or
small distances).
Comparison of eq.(9.4) with eq.(9.3) allows then to estimate the
relative values of the electric and strong charges in our
approach. In fact if we set in a first approximation $J'\simeq J$
we obtain
\begin{equation}
g\simeq 10e
\end{equation}
in good approximation with the known experimental values. Even if
this is only a preliminary indication, one could expect to find
that the strong charge $g$ is larger than the electric charge
simply because the volumes of the unit spheres decrease after a
certain dimension (since $V(B_{2k+2})/V(B_{2k})= \pi/(k+1)$  which
is $<1$ for $k>2$).

>From this preliminary indication we have then that, for purely
geometrical reasons, both masses and charges of fermions should
increase with the dimensions of the multiplets as in fact it
appears in the empirical world. This will be further analyzed
elsewhere.

\subsection{The constraint equations}

While determining underlying geometry, constraint equations might
also have some direct physical meanings which deserve further
study. One is certainly triality, arising from the one constraint
equation for 8-component spinors, which could have a role in
supersymmetry. Another is the role of the ten constraint equations
for 16-component spinors manifest in the covariant quantization of
super strings, even if at the ghost stage \cite{seventeen}.

But also the sixty-six constraint equations for 32-component
spinors correlating the baryon quadruplet $\Theta_B$ with the
lepton one $\Theta_{\cll}$ could perhaps solve the problem of the
often predicted and never found lifetime of the proton, imposing
its stability \cite{seven}.

\subsection{Laws and phenomena}

The equations of motion like the ones discussed in chapter 4:
\begin{equation}
\left[ i\frac{\partial}{\partial x_\mu} G^\mu + \sum^{2n-1}_{j=5}
P_j(x)G_j+P_{2n}(x)\right] \Psi (x) =0
\end{equation}
represent the evolution of the wavefunction $\Psi$ of a system in
space-time $M=\R^{1,3}$. For given boundary conditions a solution
$\Psi (x)$ represents the evolution of a particular phenomenon of
that system. If that solution is inserted in eq.(9.6) this becomes
an identity for all values of $x$;  meaning the validity of the
evolution of that phenomenon for the whole space and for all
times.

One can write the equation in the equivalent Cartan's form:
\begin{equation}
\left(\sum^{2n-1}_{j=1} \ P_jG^j+P_{2n}\right) \Psi (P) =0
\end{equation}
and then the existence of non null solutions implies the condition
\begin{equation}
P_\alpha P^\alpha =0,\qquad \alpha = 1,2,\dots ,2n
\end{equation}

Now setting:
\begin{equation}
P_\alpha = \widetilde\Psi ' G_\alpha\Psi '
\end{equation}
and assuming $\Psi '$ simple or pure eq.(9.8) is satisfied because
of Proposition 3, and then there are solutions of eq.(9.7) of the
form:
\begin{equation}
\Psi = \Psi (\Psi ')
\end{equation}
which, if inserted in eq.(9.7), renders it an identity in spinor
space, meaning the spinorial origin, outside of space-time, of the
law represented by the equation. The Cartan's equations for simple
spinors seem to be the only ones to present this double
possibility of becoming identities: both in space-time and in
spinor space and might present interesting epistemological
consequences which will be analyzed elsewhere.

There are more aspects of theoretical physics in which simple
spinors might play a role, specially, those concerned with null or
optical vectors, like strings and superstrings, which may be
conceived as continuous sums, or integrals, of null vectors, where
they could help us to understand also the deep reason why, for
describing elementary physical phenomena, the euclidean elementary
concept of point-event has to be rather substituted by that of
string which may be correlated with the \'E. Cartan's hypothesis
of simple spinor geometry as underlying euclidean geometry
\cite{one}, \cite{seven}.

Concluding we may affirm that simple spinors are both elegant and
rich of possibilities for the purely geometrical interpretation of
the elementary physical world. We have here merely hinted to some
of them. The resulting panorama, seems to us however, even at this
preliminary stage, to present deep and fascinating aspects
deserving further study.

\end{document}